\documentclass[letterpaper, 10pt, conference]{ieeeconf}
\IEEEoverridecommandlockouts 
\usepackage{subcaption}
\usepackage[font=small]{caption}
\usepackage{subcaption}
\usepackage{tikz}
\usepackage{xcolor}
\usepackage{graphicx}
\usepackage{amsmath,amssymb}
\usepackage[ruled,vlined]{algorithm2e}
\usepackage[utf8]{inputenc}

\let\labelindent\relax
 
\usepackage{amsthm}
\usepackage{amsfonts}
\usepackage{amsmath}
\usepackage{amssymb}

\usepackage{float}

\newtheorem{theorem}{Theorem}
\newtheorem{proposition}{Proposition}
\newtheorem{assumption}{Assumption}
\makeatletter
\let\NAT@parse\undefined
\makeatother
\usepackage{url}
\usepackage[colorlinks,citecolor=blue,linkcolor=red]{hyperref}
\usepackage{cite}
\usepackage{cleveref}
\crefrangeformat{equation}{(#3#1#4)--(#5#2#6)}
\usepackage{import}
\usepackage{enumitem}
\newtheorem{definition}{Definition}

\usepackage{booktabs}
\title{\LARGE \bf Learning Closed-Loop Parametric Nash Equilibria of Multi-Agent Collaborative Field Coverage}

\author{
    Jushan Chen and 
    Santiago Paternain
    \thanks{$^{2}$The authors are with the Department of Electrical, Computer, and Systems Engineering, Rensselaer Polytechnic Institute, 110 8th St, Troy, NY 12180, USA. {\tt\small \{chenj72, paters\}@rpi.edu}}
}

\begin{document}

\setlength{\textfloatsep}{5pt plus 2pt minus 4pt}
\setlength{\dbltextfloatsep}{3pt}
\setlist{nosep} 

\newcommand \blue[1]         {{\color{blue}#1}}
\newcommand\green[1]       {{\color[rgb]{0.10,0.50,0.10}#1}}
\newcommand \red[1]         {{\color{red}#1}}

\maketitle
\thispagestyle{empty}
\pagestyle{empty}

\begin{abstract}
Multi-agent reinforcement learning is a challenging and active field of research due to the inherent nonstationary property and coupling between agents. A popular approach to modeling the multi-agent interactions underlying the multi-agent RL problem is the Markov Game. There is a special type of Markov Game, termed Markov Potential Game, which allows us to reduce the Markov Game to a single-objective optimal control problem where the objective function is a potential function. In this work, we prove that a multi-agent collaborative field coverage problem, which is found in many engineering applications, can be formulated as a Markov Potential Game, and we can learn a parameterized closed-loop Nash Equilibrium by solving an equivalent single-objective optimal control problem. As a result, our algorithm is 10x faster during training compared to a game-theoretic baseline and converges faster during policy execution.
\end{abstract}

\section{INTRODUCTION} \label{sec:introduction}
Multi-robot collaborative field coverage refers to a problem setting where a group of autonomous agents collectively maximizes the coverage of a map. Such a problem is studied in various engineering applications. Previous works proposed solutions using methods such as Voronoi cells \cite{voronoi_1,voronoi_2} and potential fields \cite{Ge2002DynamicMP}. However, many prior works relied on full model knowledge of the environment. In reality, it is challenging to obtain an accurate world model. To tackle this challenge, reinforcement learning is a natural solution. In particular, multi-agent reinforcement learning (MARL) is becoming increasingly popular for solving problems involving interactive decision-making processes involving multiple agents \cite{Foerster2016,Zazo2016,Subramanian2019,Zhang2018, Oliehoek2016,Arslan2017,Yongacoglu2019,cui2023breaking,li2022minimax,agorio2025cooperative}. MARL is effective in various application scenarios, such as collaborative autonomy for urban mobility \cite{Enders2022HybridMD,Choudhury2022aamas}. However, a critical challenge of MARL is non-stationarity, since as agents modify their policies, the rewards and transitions potentially change over time \cite{Zhang2019MultiAgentRL}. 
\begin{figure}[t]
    \centering
    \includegraphics[scale=0.8]{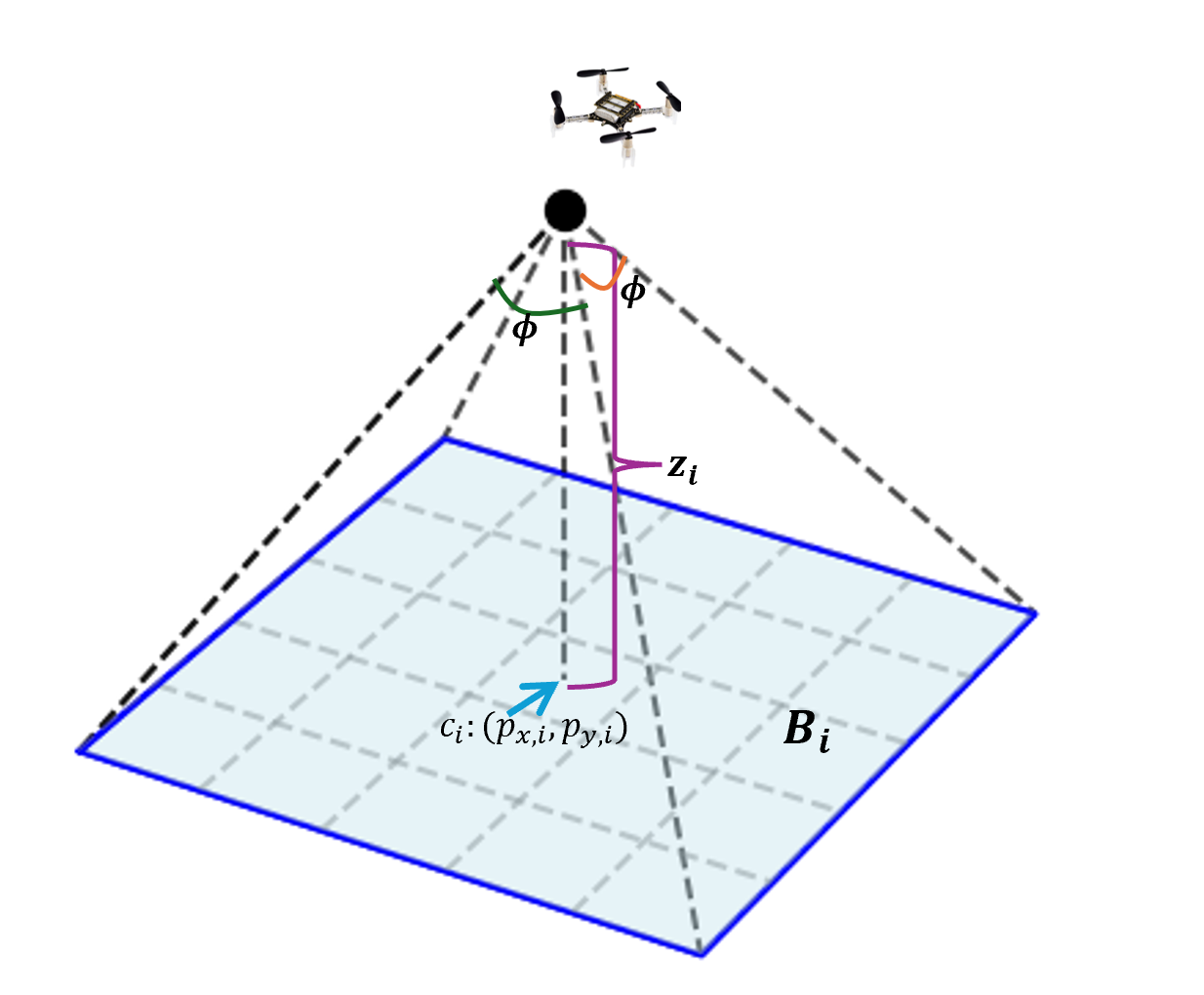}
    \caption{Illustration of the field of view (FOV) of a single agent. We assume that the agent is a UAV with small half-angles limiting its field of view. In this example, the FOV is a square. In general, the FOV has a rectangular shape.}
    \label{fig:FOV_fig}
\end{figure}


Game-theoretic formulations are often used to tackle this challenge. From a game-theoretic perspective, the multi-agent interactive decision-making process is most commonly referred to as a Markov Game \cite{Shapley1953StochasticG}. A common solution concept to finding the equilibrium state of all agents in a Markov Game is \textit{Nash Equilibrium}, which defines an equilibrium strategy from which none of the agents have any incentive to deviate \cite{Baar1982DynamicNG}. One special structure of dynamic games is potential games. Potential game is a general framework to model cooperation between agents \cite{Monderer1994PotentialG}, where there is some potential function shared by all agents such that if any agent unilaterally changes its strategy, the total change in its reward is equivalent to or proportional to the total change in the potential function. Several works have studied the theoretical properties and applications of potential games \cite{maulik_game, mpg_energy_resource,mpg_paper,dp-ilqr, imagined_potential, GonzlezSnchez2013DiscretetimeSC, Scutari2006PotentialGA, Zazo2015DynamicPG, Kavuncu2021PotentialIA}, showing effectiveness in solving trajectory planning, resource allocation, etc. A Markov extension to potential games, i.e., Markov Potential Games, is studied in \cite{mpg_paper}. Recently, \cite{cooperative_cover} leveraged game-theoretic concepts to tackle the multi-robot field coverage problem. In particular, they used the Correlated Equilibrium (CE) as a solution strategy and employed Q-learning to find an equilibrium. However, their algorithm is barely scalable beyond 2 agents due to the complexity of solving the underlying optimization problem arising from CE. In contrast, in this work, we propose to use the Nash Equilibrium (NE) as the solution strategy, and we leverage a Markov Potential Game approach to achieve a more efficient and scalable algorithm for finding the equilibrium.
Our contributions to this work are twofold:
\begin{itemize}
    \item We prove that the multi-robot collaborative field coverage problem can be modeled as a Markov Potential Game.
    \item We show that by leveraging the Markov Potential Game, our approach is more scalable and 10x faster during training than the baseline.
\end{itemize}

We organize this work as follows: in Sec. \ref{sec:problem-formulation}, we formulate the multi-robot collaborative field coverage problem as a parametric Markov Game; In Sec. \ref{sec:mpg_algorithm}, we show that the underlying parametric Markov Game is in fact a Markov Potential Game and propose an efficient algorithm to find the closed-loop parametric NE; In Sec. \ref{sec:simulation_studies}, we present simulations studies showcasing the superiority of our algorithm in terms of scalability and training efficiently, in comparison to a baseline solver. We conclude this work in Sec. \ref{sec:conclusion}.


\section{Problem Formulation} \label{sec:problem-formulation}
\subsection{Multi-Agent Collaborative Field Coverage}
\begin{figure}[t]
    \centering
    \includegraphics[scale=0.8]{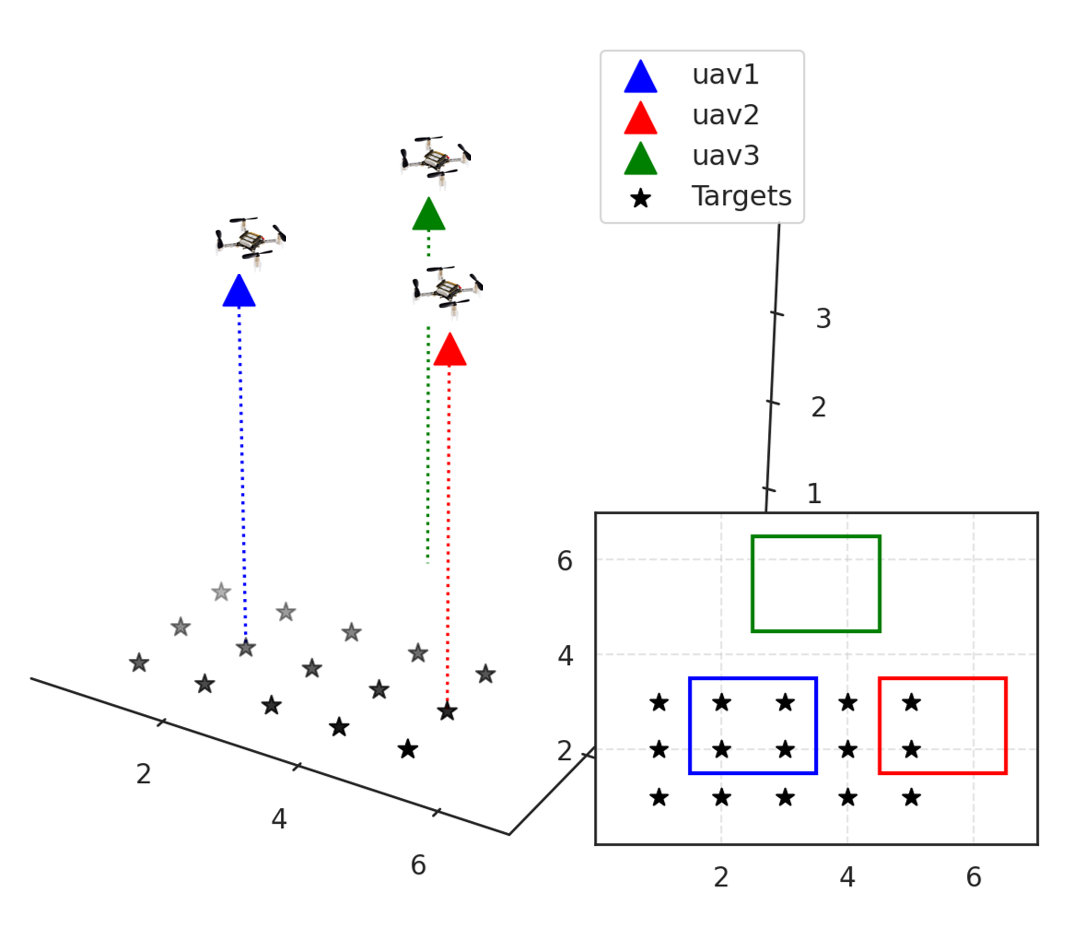}
    \caption{A simplified visualization of multiple UAVs attempting to cover a set of targets. Each UAV has a limited FOV shown as a bounding box. The static targets are shown as stars colored in black.}
    \label{fig:env_render}
\end{figure}

This work considers a team of $N$ agents (or robots) denoted by  $\mathcal{N}=\left\{1,\ldots,N\right\}$ covering a planar area of interest. Each agent's state is denoted by $s_i \in \mathbb{R}^3$ representing its position $s_i = [{p_{x_i}},{p_{y_i}},{p_{z_i}}]$ for all $i\in\mathcal{N}$. 
At each time step $t\in \mathbb{N}_0$ the joint state of the system is denoted by $s_t =  (s_{1,t}, s_{2,t}, \dots, s_{N,t}) \in \mathcal{S}$. Similarly, let each agent's action be a velocity $a_i \in \mathbb{R}^3$ and denote the joint action of the team at time $t$ by $a_t = (a_{1,t}, a_{2,t}, \dots, a_{N,t}) \in \mathcal{A}$.
The individual state transition function is defined such that $P_i(s_{i,t+1}|s_{i,t+1},a_{i,t})$ refers to the likelihood of agent $i$'s state $s_{i,t}$ transitioning to a new state $s_{i,t+1}$ given that an action $a_{i,t}\in \mathcal{A}_i$ is taken. 
For example, the transition can correspond to integrating the agent's velocity, where a low-level controller tracks the setpoint velocity $a_{i,t}$. We assume the low-level controller can also take care of collision avoidance. Such schemes are common in the literature~\cite{cooperative_cover, Westheider2023MultiUAVAP}. We define a field of interest (FOI) by a set $F\subset \mathbb{R}^2$ containing some static targets to be covered, which has an arbitrary shape. Each agent $i$ has a set of fixed half-angles associated with its downward-pointing camera, denoted by $\phi$ as shown in Fig. \ref{fig:FOV_fig}. We denote by ${c_i}:=(p_{x_{i}},p_{y_{i}}) \in \mathbb{R}^2$ the horizontal components of the position of agent $i$. Given a certain altitude of the agent $p_{z,i} \in \mathbb{R}$, the set of targets captured in agent $i$'s field of view (FOV) is given by the following expression \cite{cooperative_cover}

\begin{equation}
B_i= \left\{q\in F \mid \frac{\left\|q-c_i\right\|}{p_{z_i}} \leq \tan (\phi)\right\}.
\label{eqn:fov_coverage}
\end{equation}
The goal of the team of $N$ agents is to maximize their collective coverage of the targets while minimizing their mutual overlaps, as illustrated in Fig. \ref{fig:env_render}. From a purely geometric perspective, we ignore the time index and denote the area covered by agent $i$ as ${f_i(s_i)} = \int_{B_{i}} \mathbf{1}dq$, and denote its overlap with another agent $j$ as $O_{ij}(s_i,s_j) = \int_{q \in B_i \cap B_j}\mathbf{1}dq$.
The net coverage area of each agent $i\in\mathcal{N}$ is given by:
\begin{equation}
\begin{aligned}
r_i & =f_i(s_i)-\sum_{j\in\mathcal{N},j\neq i}O_{ij}(s_i, s_j) \\&=\int_{q\in B_{i}} \mathbf{1}dq-\sum_{j\in\mathcal{N},j\neq i}\int_{q \in B_i \cap B_j}\mathbf{1}dq.
\end{aligned}
\label{eqn:agent_i_objective}
\end{equation}
By reformulating each agent's net coverage as a reward function, we can formulate the problem of maximizing collective coverage as a Markov Game. We discuss relevant details in the following sections.

\subsection{Parametric Closed-Loop NE of a Markov Game}
\begin{figure*}[t]
    \centering
    \includegraphics[scale = 0.7]{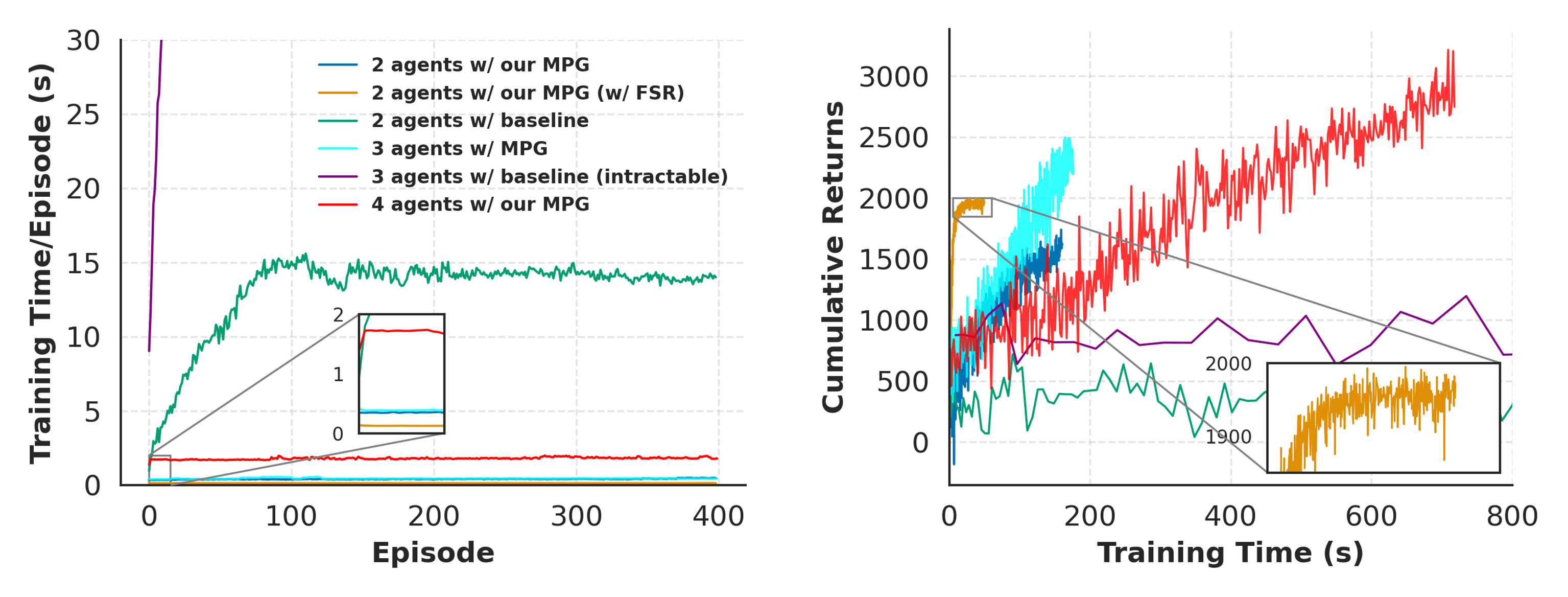}
    \caption{On the left is a comparison of the training time between our algorithm and a game-theoretic baseline \cite{cooperative_cover}, and on the right is a comparison of cumulative returns versus training time between our algorithm and the baseline. We run each algorithm for 400 episodes with 200 steps per episode. To make a fair comparison, we add an additional baseline (orange colored) by parameterizing the $Q$ function with the FSR used by the baseline.}
    \label{fig:training_time}
\end{figure*}

Let $w_i \in \mathbb{W}_i \subseteq \mathbb{R}^{\mathcal{W}_i}$ denote agent $i$'s parameter vector of length $\mathcal{W}_i$ for the parameterized policy $\pi_i$. This parameterization defines a set of policies given by
\[
\Omega^w_i \triangleq \{ \pi_i(\cdot, w_i) : w_i \in \mathbb{W}_i \}.
\]
We define joint policy space as the product of local policy spaces:
\begin{align}
    \Omega^w &\triangleq \prod_{i \in \mathcal{N}} \Omega^w_i, \quad
    \mathbb{W} \triangleq \prod_{i \in \mathcal{N}} \mathbb{W}_i,  \quad \mathcal{W} \triangleq \sum_{i \in \mathcal{N}} \mathcal{W}_i,
\end{align}
We also denote the set of parametric policies, $\Omega^w$, as a finite-dimensional function space with parameter $w \in \mathbb{W} \subseteq \mathbb{R}^{\mathcal{W}}$: $\Omega^w \triangleq \{ \pi(\cdot, w) : w \in \mathbb{W]} \}$. For a given $w$, the parametric policy is a mapping from states to actions, i.e., $\pi(\cdot, w) : \mathcal{S} \to \mathcal{A}$. 
Furthermore, we denote $w_{-i}$ as the parameters of all agents \textit{except agent $i$}, so that we can also equivalently define:
\begin{align}
    w &= (w_i, w_{-i}), \nonumber \\
    \pi(\cdot, w) &= \pi(\cdot, (w_i, w_{-i})) \nonumber \\
    &= (\pi_i(\cdot, w_i), \pi_{-i}(\cdot, w_{-i})).
\end{align}
We now define a \textit{parametric Markov Game} as follows:
\begin{equation}
\mathcal{G} :
\begin{cases}
\displaystyle \max_{w_i \in \mathbb{W}_i} \mathbb{E}\left[
            \sum_{t=0}^{\infty} \gamma^t r_{i} 
            \left( \mathbf{s}_t, \pi_i(\mathbf{s}_t, w_i), 
            \pi_{-i}(\mathbf{s}_t, w_{-i})\right)
        \right] \\
\text{s.t.} \quad s_{t+1} = P(s_t, \pi(s_t, (w_i,w_{-i})), \\[5pt]
\forall i \in \mathcal{N} 
\end{cases}
\label{eqn:parametric_markov}
\end{equation}
where $r_{i}$ now refers to the reward. Eqn~\eqref{eqn:parametric_markov} describes the underlying decision-making process of all agents over time, and at each time $t$ the joint state $s_t$ transitions into the next joint state $s_{t+1}$ resulting from action $\pi(s_t,(w_i,w_{-i}))$. The objective of each agent $i$ is to maximize its cumulative discounted rewards by learning a policy parameterized by $w_i$. Before we formally define the Nash Equilibrium, we consider the following assumptions.
\begin{assumption}\label{assumption_convex}
    $\mathcal{S}$ and $\mathbb{W}$ are nonempty and convex.
\end{assumption}
\begin{assumption}\label{assumption_}
   The state transition function $P$ is continuously differentiable.
\end{assumption}
\begin{assumption}\label{assumption_bounded}
   The rewards satisfy $\mathbb{E}[r_i(s_t,a_t)]^\infty_t > -\infty$ for at least one $(s_t,a_t)$, and $\mathbb{E}[r_i(s_t,a_t)]^\infty_t < \infty$ for all $(x_t,a_t)$, $t \in [0,\dots,\infty]$.
\end{assumption}
Assumptions \ref{assumption_convex} and \ref{assumption_} are mild and Assumption \ref{assumption_bounded} guarantees the existence of a \textit{parametric closed-loop Nash Equilibrium (PCL-NE)}\cite{mpg_paper}, which we define next. 
\begin{definition}
A \textit{parametric closed-loop Nash equilibrium (PCL-NE)} of $\mathcal{G}$~\eqref{eqn:parametric_markov} is a vector 
$w^\star = (w_i^\star, w_{-i}^\star) \in \mathbb{R}^{\mathcal{W}}$ that satisfies:
\begin{equation}
    \begin{split}
        \mathbb{E} \left[
            \sum_{t=0}^{\infty} \gamma^t r_i 
            \left( \mathbf{s}_t, \pi_i(\mathbf{s}_t, w_i^\star), 
            \pi_{-i}(\mathbf{s}_t, w_{-i}^\star)\right)
        \right] \\
        \geq \mathbb{E} \left[
            \sum_{i=0}^{\infty} \gamma^t r_i 
            \left( \mathbf{s}_t, \pi_i(\mathbf{s}_t, w_i), 
            \pi_{-i}(\mathbf{s}_t, w_{-i}^\star) \right)
        \right],
        \forall i \in \mathcal{N}.
    \end{split}
\end{equation}
\end{definition}
\vspace{-5pt}
\noindent Intuitively, the PCL-NE is defined as the parametric joint policy $\pi(\cdot|w^*)$ from which if an agent $i$ unilaterally deviates, no benefit or any gain in rewards is possible for all agents $i$. Thus, no agents would have any incentive to unilaterally deviate from such equilibrium strategy $\pi(\cdot|w^*)$. However, finding the PCL-NE is challenging, as we have a set of coupled optimal control problems in the Markov Game. As discussed in \cite{mpg_paper}, the drawback of the standard approach is that we first need to guess some parametric policies from the function space $\Omega$ and then check if they satisfy the optimality conditions of the problem. A method that leverages CE was proposed in \cite{cooperative_cover} to solve the multi-agent collaborative field overage problem. Yet, it does not scale (see Figure \ref{fig:training_time}). To tackle such challenges, we propose to leverage \textbf{Markov Potential Game (MPG)} leading to a more efficient solution strategy.

\section{Multi-agent Collaborative Field Coverage as a Markov Potential Game} \label{sec:mpg_algorithm}
\begin{figure*}[t]
    \centering
    \includegraphics[width=\linewidth]{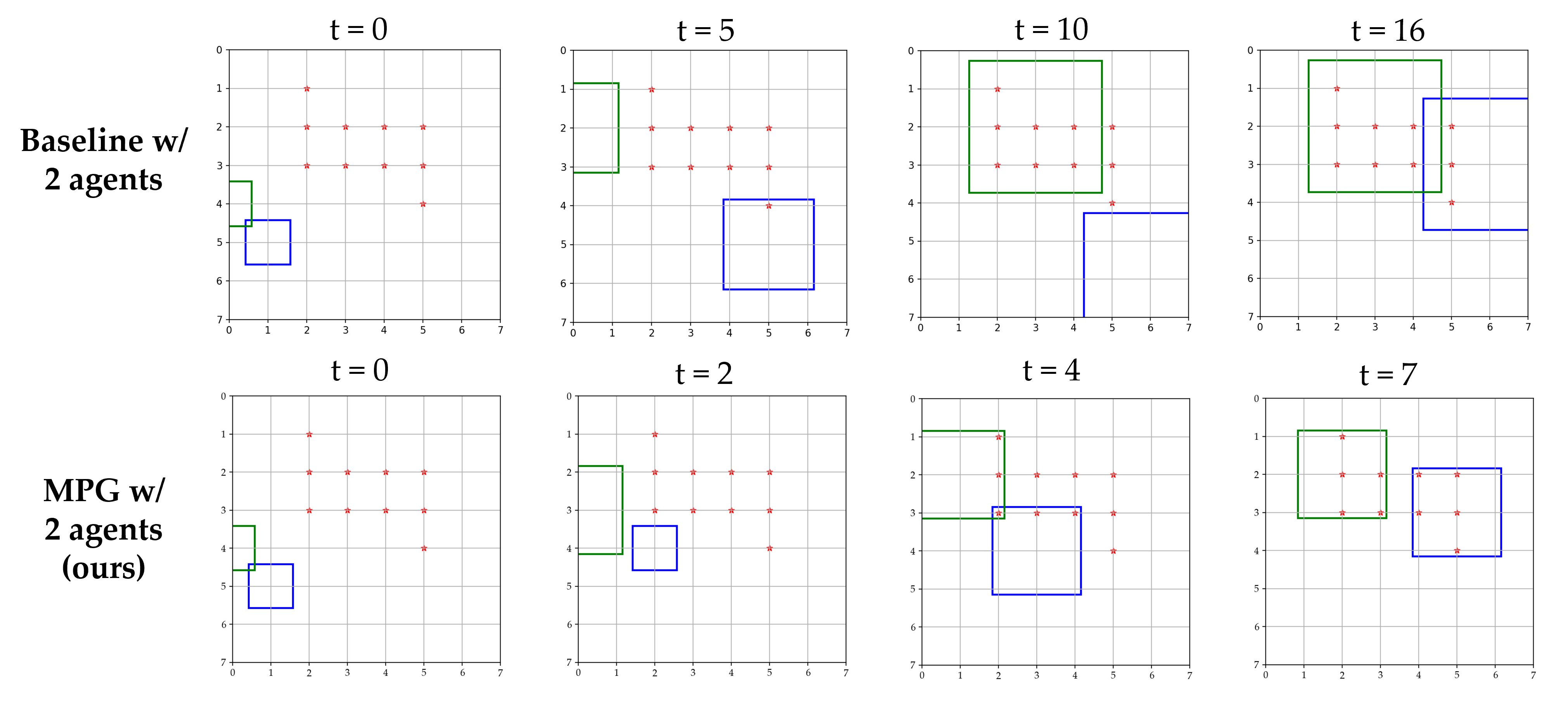}
    \caption{Policy execution to recover a PCL-NE for the 2-agent scenario. We set the maximum number of transition steps to 20. We observe that Algorithm . \ref{alg:dqn_potential} converges to an equilibrium much faster than the baseline.}
    \label{fig:2_agent_test_coverage}
\end{figure*}
\begin{algorithm}[t]
\SetKwInOut{Input}{Inputs}
\caption{\textbf{Q-Learning for PCL-NE via MPG}}
\label{alg:dqn_potential}
\BlankLine
\BlankLine
\textbf{Initialize:} DNN parameterized by $\zeta$, $stepCount = 0$, $\epsilon_0$ = 10000, $T_{max}=200$,
\BlankLine
\For{\(\text{episode} \gets 0 \,\KwTo\, K_{\text{episodes}}\)}{
  \(s \gets \text{ResetEnvironment}()\)\;
  \(R_{\text{episode}} \gets 0\)\;
  \For{\(t \gets 0 \,\KwTo\, T_{\max}-1\)}{
    \(\displaystyle \varepsilon \;\gets\;
       \max\!\Bigl(\varepsilon_{\min},\;\;
                   \varepsilon_{\max}\,\exp\!\bigl(-\,\tfrac{stepCount}{\epsilon_0}\bigr)\Bigr)
    \)\;
    \uIf{\(\text{UniformRand}(0,1) < \varepsilon\)}{
       \(\pi(s|w) \gets \text{RandomAction}\)\; 
    }
    \Else{
       \(\pi(s|w) \gets \arg\max_{a'} \;Q_{\zeta}(s,\,a')\)\;
    }
    \(s' \gets \text{NextState}(s,\,a)\)\;
    \(\displaystyle r \;\gets\; J_{t}\)\;
    \(R_{\text{episode}} \gets R_{\text{episode}} + r\)\;
    $stepCount \gets stepCount + 1$\;
    \(\displaystyle \delta \;\gets\; r + \gamma \,\max_{a'}\Bigl[Q_{\zeta}(s',\,a')\Bigr]
                         - Q_{\zeta}(s,\,a)\)\;
    \text{Perform gradient descent to minimize } \(||\delta||^2\) and backpropagate $\zeta$\;
    \(s \gets s'\)\;
    \If{\(stepCount\geq T_{max}-1\)}{
      \textbf{break}\;
    }
  }
  \text{Record}\(\bigl(R_{\text{episode}}\bigr)\)\;
}
\Return {$\zeta^\star$ and policy $\pi(s|w^*)$}
\end{algorithm}
An MPG is a special instance of a Markov game. The benefit of leveraging an MPG is that we can reduce the set of coupled optimal control problems (OCPs) into an equivalent single-objective OCP, thereby greatly simplifying the solution process. We denote $s^r_{i,t}$ as the components of the joint state vector that influence the reward of agent $i$ directly. Similarly, we denote $s^\pi_{i,t}$ as the components of the joint state vector that influence the policy of agent $i$ directly. In addition, we denote the union of state components that influence the policy of agent $i$ or influence the reward of agent $i$ as $\tilde{s}_{i,t}$. We formally define the MPG according to the following theorem \cite{mpg_paper}.
\begin{theorem}
Let us consider the Markov Game $\mathcal{G}$ defined in ~\eqref{eqn:parametric_markov} let Assumptions \ref{assumption_convex}-\ref{assumption_bounded} hold. In addition, we assume that the reward of each agent $i$, $r_{i}$ is twice continuously differentiable in $\mathcal{S} \times \mathcal{A}$. Then, $\mathcal{G}$ is an MPG if and only if:
\noindent 1) \textit{the reward function of every agent $i$ can be expressed as the sum of a term $J(\cdot)$ common to all agents plus another term $\Theta_i(\cdot)$ that depends neither on its own state-component vector nor on its policy parameter:}
\begin{equation}
    \begin{aligned}
        r_{i}\big(&s^r_{i,t}, \pi_i(s^\pi_{i,t}, w_i), \pi_{-i}(s^\pi_{-i,t}, w_{-i})\big) \\
        &= J\left(s_t, \pi(s_t, w)\right) \\
        &\quad + \Theta_i\left(s^r_{-i,t}, \pi_{-i}(s^\pi_{-i,t}, w_{-i})\right), \quad \forall i \in \mathcal{N};
    \end{aligned}
\end{equation}
and 2) \textit{the following condition on the non-common term $\Theta_i$ holds:}
\begin{equation}
    \mathbb{E} \left[ \nabla_{\tilde{s}_{i,t}} \Theta_i\left(s^r_{-i,t}, \pi_{-i}(s^\pi_{-i,t}, w_{-i})\right) \right] = 0.
    \label{eqn:mpg_condition2}
\end{equation}
\label{theorem:potential_MPG}
\end{theorem}
\vspace{-6mm}
Intuitively, the potential function $J(\cdot)$ is common to all agents' states and policies, whereas $\Theta_i(\cdot)$ must not depend on the state components or policy parameters of agent $i$ itself. In other words, the term $\Theta_i(\cdot)$ is used to balance any ``asymmetry", such that if agent $i$ unilaterally deviates from its policy, then the associated change in its reward $\Delta r_i(\cdot)$ must be captured by the change in the potential $\Delta J(\cdot)$ completely. Most importantly, the advantage of leveraging MPG follows that once we have an MPG with a potential function $J(\cdot)$, solving the original parametric Markov Game~\eqref{eqn:parametric_markov} is reduced to an equivalent \textit{single-objective OCP}\cite{mpg_paper}:
{\small
\begin{equation}
\mathcal{G}_{MPG} : 
\begin{cases}
\displaystyle \max_{w \in \mathbb{W}} \mathbb{E}\left[ \sum_{t=0}^{\infty} \gamma^t J (s_t, \pi(s_t,w)) \right] \\[10pt]
\text{s.t.} \quad s_{t+1} = P(s_t, \pi(s_t, w)), \\[5pt]
\end{cases}
\label{eqn:parametric_MPG}
\end{equation}
}
\noindent We now claim that our multi-agent collaborative field coverage problem is indeed an MPG. 
\begin{proposition}
The multi-agent collaborative field coverage problem is an MPG, with a potential function
\begin{equation}
J(q, s_t,\pi(s_t,w)) = \sum^\mathcal{N}_{i} f_{i}(q,s_{i,t}) \allowbreak - \sum_{\substack{j\neq i,\\ j\in \mathcal{N}}} O_{ij}(q, s_{i,t},s_{j,t}),
\end{equation}
where $f_{i}(q,s_{i,t})$ denotes the total coverage area by agent $i$ at time $t$, and $\sum_{j\neq i, j\in \mathcal{N}} O_{ij}(q, s_{i,t},s_{j,t})$ denotes the overlap between unique pairs of agents $(i,j)$ at time $t$.
\label{prop:mpg}
\end{proposition}
\begin{proof}
We prove Proposition. \ref{prop:mpg} following Theorem. \ref{theorem:potential_MPG}. Upon inspection of~\eqref{eqn:fov_coverage}, we define $r_{i}(\cdot)$, the reward for agent $i$ at time $t$ as the following expression:
\begin{equation}
\begin{split}
    r_{i}(q,s^{r}_{i,t},\pi_i(s^{\pi}_{i,t},w_i),\pi_{-i}(s^{\pi}_{-i,t},w_{-i}))& \\= f_{i}(q,s_{i,t})-\sum_{j \neq i, j\in \mathcal{N}} O_{ij}(q, s_{i,t},s_{j,t}),
\label{eqn:reward_proof}
\end{split}
\end{equation}
\begin{figure*}[t]
    \centering
    \includegraphics[width=\linewidth]{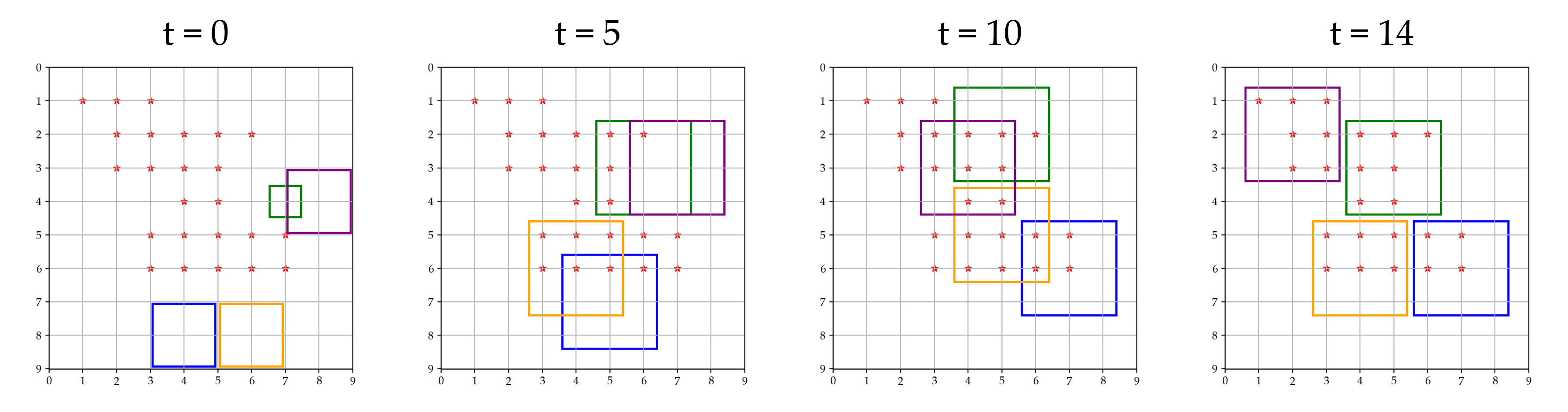}
    \caption{Policy execution for the 4-agent scenario with Algorithm. \ref{alg:dqn_potential}. } 
    \label{fig:test_4_agent_covearge}
\end{figure*}
\vspace{-4pt}
Upon inspection of Eqn~\eqref{eqn:reward_proof} we can recover the Markov Potential game structure according to Theorem. \ref{theorem:potential_MPG}. We let
\begin{equation}
J(s_t,\pi(s_t,w))= \sum_{i=1}^N f_{i}(q,s_{i,t})-\sum_{j\neq i, j \in \mathcal{N}}O_{ij}(q, s_{i,t},s_{j,t}),
\label{eqn:MPG_proof}
\end{equation}
and let
\begin{equation}
\begin{aligned}
    \Theta_i(s^{r}_{-i,t}) = - \sum_{j\neq i, j\in \mathcal{N}} f_{j,t}(q,s_{j,t}),
\end{aligned}
\label{eqn:mpg_proof2}
\end{equation}
where $J(\cdot)$~\eqref{eqn:MPG_proof} is common to all agents and is equivalent to the sum of local coverage from all agents subtracting all pairwise overlaps, whereas the term $\Theta_i(\cdot)$~\eqref{eqn:mpg_proof2} neither depends on the state components of agent $i$ nor on the policy parameter of agent $i$. Agent $i$'s original reward $r_{i}(\cdot)$ is then recovered by $r_{i} = J(s_t,\pi(s_t,w)) + \Theta_i(s^r_{-i,t})$, where the terms on the right-hand side are given in Eqn~\eqref{eqn:MPG_proof} and~\eqref{eqn:mpg_proof2} respectively. Upon inspection, the condition in~\eqref{eqn:mpg_condition2} holds in this case
\label{proof:mpg}
\end{proof}

To learn the PCL-NE to our MPG, we use the DQN \cite{Mnih2013PlayingAW} algorithm. Our training procedure is outlined in Algorithm. \ref{alg:dqn_potential}. We first initialize a two-layer deep neural network (DNN) with weights $\zeta$ as the $Q$-function approximator, where $Q_\zeta(s_t,a_t)$ takes the state and action pair $(s_t,a_t)$ as input and returns a single predicted state-action value as output. We employ an $\epsilon$-greedy strategy for action selection with an exponential decay schedule. At each $t$, the immediate global reward $r_t$ is taken as the current potential function $J_t$. After training is complete, we use the standard iterative best response method for decentralized policy execution.

\section{Simulation Studies} \label{sec:simulation_studies}
\subsection{Simulations Settings}
\label{sec:simulation_setting}
We present two sets of simulation scenarios. In the first scenario, we deploy 2 UAVs and set up a 3-D grid world with dimensions $(7 \times 7 \times 4) meters^3$, where 4 $meters$ is the maximum height; in the second scenario, we deploy 4 UAVs and set up a grid world with dimensions $(9 \times 9 \times 4) meters^3$. In both cases, we discretize the grid world in cells of size $1\times 1\times 1$. We assume a kinematic model for each agent, and we assign 6 possible actions to each agent (move north, move south, move left, move right, move up, and move down) such that its state transition is given by:
\begin{equation}
    s_{i,t+1} = s_{i,t} + a_{i,t}, \forall i\in \mathcal{N}.
\end{equation}
For each scenario, we randomly generate a set of target locations to form an FOI with a shape unknown to the agents. At each environment state transition time $t$, we choose the immediate reward $r_t(\cdot)$ as the current potential function $J(q,s_t,\pi(s_t,w))$, which is previously defined in~\eqref{eqn:MPG_proof}.
We solve our Markov Potential Game according to the procedures outlined in Alg. \ref{alg:dqn_potential}. We first initialize a two-layer DNN $Q_\zeta(s,a)$ that takes the joint state $s$ and joint action $a$ as input and outputs a single $Q$ value. We adopt an $\epsilon$-greedy strategy, and with a probability greater than $1-\epsilon$, the action is taken as the $argmax$ over the predicted Q value. DNN parameters $\zeta$ are backpropagated via stochastic gradient descent on the TD-error $||\delta||^2$. We choose a hidden size of 64 for each layer of the DNN. We run the training loop for 400 episodes, in which each episode has a maximum of 200 state transition steps. We choose a learning rate of $\alpha = 10^{-3}$ for SGD, a discount factor $\gamma = 0.9$, and a batch size of 64 for the replay buffer for $Q$-learning. We assume that the FOV half-angles $\phi^{\intercal}$ = $[30^\circ, 30^\circ]^\intercal$. We run our algorithm~\eqref{alg:dqn_potential} and the baseline on a PC with an RTX 4070 GPU with 12 gigabytes of RAM.
\subsection{Simulation Results}
We compare our algorithm against the baseline in \cite{cooperative_cover}. We evaluate our algorithm with $\mathcal{N}$ =  2,3,4 agents, respectively. The baseline is barely tractable beyond 2 agents, and thus, we only run our algorithm on the 4-agent scenario. Since the baseline \cite{cooperative_cover} uses the Fixed Sparse Representation (FSR) scheme to parameterize the $Q$ function, for a fair comparison, we also run our algorithm for the 2-agent scenario with the same FSR parameterization scheme. From Fig. \ref{fig:training_time}, we observe that the training time of the baseline algorithm for the 2-agent scenario gradually increases and peaks at around 100 episodes, which is over \textbf{10x} slower than our MPG solver. We also observe that our algorithm cumulates the rewards faster than the baseline for each simulation scenario. 

We note that the baseline is not scalable beyond 2 agents: the training time increases to over 120 seconds per episode, which is clipped on the y-axis. Although both the baseline and our algorithm deploy the same $\epsilon$ decay schedule as part of the $Q$ learning process, the baseline becomes less and less tractable due to the scalability issues arising from finding the underlying CE. We refer the readers to \cite{cooperative_cover} for more implementation details regarding the baseline.

\begin{figure}
    \centering
    \includegraphics[scale=0.7]{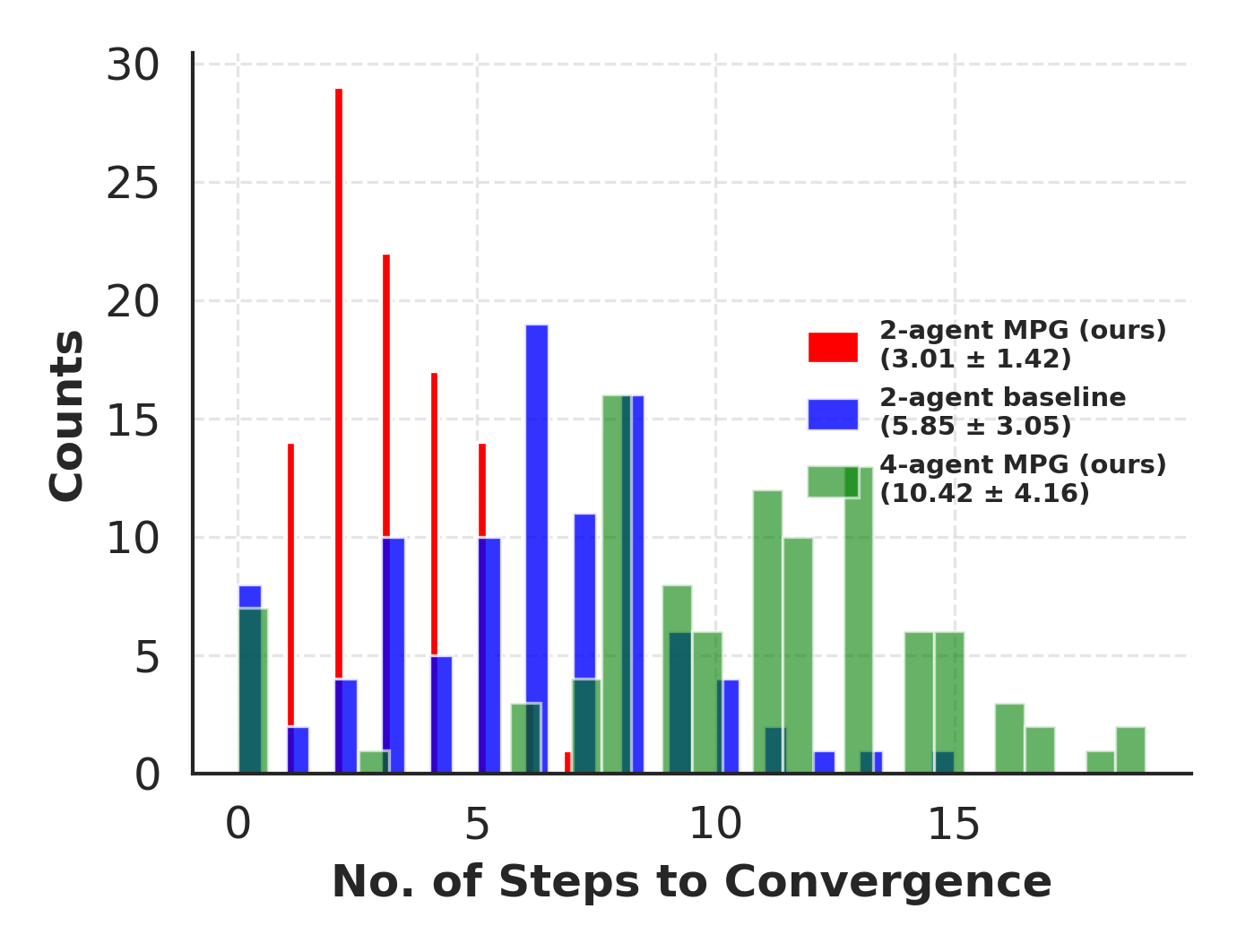}
    \caption{A comparison of the number of steps required before convergence between our algorithm and the baseline. Since the baseline is not scalable beyond 2 agents, we run the 4-agent policy using our algorithm only.}
    \label{fig:duration_statistics}
\end{figure}

We show the result for policy execution for the 2-agent scenario in Fig. \ref{fig:2_agent_test_coverage}. We observe that during policy execution our algorithm converges faster than the baseline. Since the baseline is not tractable beyond 2 agents, we only run the policy execution for the 4-agent scenario with our algorithm, as shown in Fig. \ref{fig:test_4_agent_covearge}. In addition, we run a Monte Carlo simulation for policy execution for the 2-agent scenario (baseline versus ours) and the 4-agent scenario (ours only). Upon inspection of Fig. \ref{fig:duration_statistics}, we observe that for the 2-agent scenario, the number of steps until convergence for our algorithm (red colored bars) is, on average, much fewer compared to the baseline (blue colored bars), indicating faster convergence. The associated means and standard deviations are reported in Fig. \ref{fig:duration_statistics}.

\section{Conclusion}\label{sec:conclusion}
\textbf{Summary.} We have established that the multi-agent field coverage problem can be modeled as an MPG. We then found the parameterized PCL-NE to the MPG by solving an equivalent single-objective OCP associated with a potential function. We have shown that our algorithm is several times faster to train compared to the baseline, and it converges to the optimal configuration faster than the baseline during deployment. 

\textbf{Limitations.} For future research, we would like to investigate methods to scale up our algorithm to even more agents during training, e.g., on the scale of dozens or even hundreds while still preserving tractable computation time. We would also like to study how our algorithm can be adapted to a decentralized partially observable Markov decision process instead of a fully observable Markov decision process.

\bibliographystyle{ieeetr}
\bibliography{references}

\end{document}